\documentstyle[12pt,epsfig,amssymb]{article}  

\newcommand{\ep}{\varepsilon}

\newcommand{\om}{\omega}

\newcommand{\beq}{\begin{equation}}
\newcommand{\eeq}{\end{equation}}
\newcommand{\ba}{\begin{array}}
\newcommand{\bea}{\begin{eqnarray}}
\newcommand{\ea}{\end{array}}
\newcommand{\eea}{\end{eqnarray}}

\newcommand{\bp}{{\bf p}}

\newcommand{\C}{{\cal C}}

\newcommand{\bra}{\langle}
\newcommand{\ket}{\rangle}

\newcommand{\vecp}{{\mathbf p}}

\newcommand{\vecx}{{\mathbf x}}
\newcommand{\vecy}{{\mathbf y}}
\newcommand{\vecz}{{\mathbf z}}

\newcommand{\be}{\begin{equation}}
\newcommand{\ee}{\end{equation}}
\newcommand{\bean}{\begin{eqnarray*}}
\newcommand{\eean}{\end{eqnarray*}}

\newcommand{\gm}{\gamma}
\newcommand{\Gm}{\Gamma}
\newcommand{\half}{\frac{1}{2}}

\begin{document}

\title{
\vskip -100pt
{
\begin{normalsize}
\mbox{} \hfill \\
\mbox{} \hfill HD-THEP-01-11\\
\mbox{} \hfill hep-ph/0103049\\
\mbox{} \hfill March 2001\\
\vskip  70pt
\end{normalsize} 
}
Nonequilibrium time evolution of the \\
spectral function in quantum field theory}
\author{
Gert Aarts\thanks{email: aarts@thphys.uni-heidelberg.de}
\addtocounter{footnote}{1}      
and J\"urgen Berges\thanks{email: j.berges@thphys.uni-heidelberg.de}
\addtocounter{footnote}{2}       
\\[1.ex]
{\normalsize Institut f{\"u}r Theoretische Physik}\\
{\normalsize Philosophenweg 16, 69120 Heidelberg, Germany}
}

\date{}

\maketitle 
\begin{abstract}

Transport or kinetic equations are often derived assuming a quasi-particle
(on-shell) representation of the spectral function. We investigate this
assumption using a three-loop approximation of the $2PI$ effective action
in real time, without a gradient expansion or on-shell approximation. For 
a scalar field in $1+1$ dimensions the nonlinear evolution, including the 
integration over memory kernels, can be solved numerically. 
We find that a spectral function approximately described by a nonzero 
width emerges dynamically. During the nonequilibrium time evolution the 
Wigner transformed spectral function is slowly varying, even in presence 
of strong qualitative changes in the effective particle distribution.
These results may be used to make further analytical progress towards a
quantum Boltzmann equation including off-shell effects and a nonzero width.  

\vspace{1cm}

\noindent
PACS numbers: 11.10.Wx, 05.60.Gg, 05.70.Ln 

\end{abstract}

\newpage


\section{Introduction}

Current and upcoming heavy-ion experiments at RHIC and LHC have been an
important motivation for an extensive theoretical study of kinetic or
transport equations beyond the level of the classical Boltzmann equation.
In a (nonequilibrium) environment the quasiparticle structure of the
theory can change and for instance the presence of nuclear matter in
heavy-ion collisions may affect the mass and width of vector mesons (see
e.g.\ \cite{Stachel:1999rc} for a discussion of the broadening of the
$\rho$ meson). In order to give a consistent treatment these effects
have to be incorporated in a kinetic description. Also in other areas of
physics an extension of the classical Boltzmann equation to a more
general quantum Boltzmann equation is needed. The theory of baryo- and
leptogenesis requires the solution of a large set of Boltzmann equations
with a proper inclusion of CP violating and off-shell effects in an
expanding universe \cite{Buchmuller:2000as}. In Ref.~\cite{Stoof} a
description of the dynamics of Bose-Einstein condensates using a quantum
Boltzmann equation can be found.

It is well known 
\cite{KadanoffBaym,Calzetta:1988cq,Mrowczynski:1990bu,Blaizot:2001nr} 
that a
consistent setup of kinetic equations derived from a microscopic quantum
field theory needs to incorporate two quite different aspects: the
statistical information typically described in terms of one-particle
distribution functions on the one hand, and the kinetic information about
the spectrum of the theory, encoded in the spectral function, on the other
hand.  There is a possibility to independently approximate the statistical and
the spectral aspects of the time evolution equations. In the derivation
of the classical Boltzmann equation one neglects all nontrivial kinetic
aspects of the spectral function, which is approximated by an on-shell
\mbox{$\delta$-function}. A first-order gradient expansion is then
sufficient to derive the Boltzmann equation, describing the evolution of
the statistical information only, from the underlying field theory
\cite{KadanoffBaym,Calzetta:1988cq,Mrowczynski:1990bu,Blaizot:2001nr}.  
It turns out that relaxing the
assumption of on-shell quasi-particles makes the problem considerably more
complicated. Indeed, an analysis of a self-consistent inclusion of a
nonzero width has been taken up only relatively recently
\cite{Ivanov:2000tj}. Similarly, to go beyond leading order in a gradient
expansion seems to be rather cumbersome~\cite{Joyce:2000uf}.

Recently, a successful numerical solution of the nonequilibrium time
evolution without the use of a gradient expansion or on-shell
approximation has been presented for a scalar field theory in $1+1$
dimensions \cite{Berges:2000ur}.  Both the early-time behavior and the
large-time physics of thermalization can be described remarkably well for
quantum fields \cite{Berges:2000ur}, as well as for the classical field
limit put to the test by Monte Carlo methods and numerical integration
\cite{AartsBerges:2001}. The evolution equations can be obtained from the
three-loop expansion of the closed time path generating functional for
two-particle irreducible ($2PI$) Green's functions
\cite{Calzetta:1988cq,Berges:2000ur} and comprise the classical Boltzmann
equation 
\cite{KadanoffBaym,Calzetta:1988cq,Mrowczynski:1990bu,Blaizot:2001nr}.  
The full solution of
these equations allows to directly investigate the influence of off-shell
effects on the time evolution or the validity of a gradient expansion.

In this paper we focus on the kinetic aspects of the evolution equations
and compute the nonequilibrium time dependence of the spectral function
without further approximation. The paper is organized as follows. In the
following section we briefly review the Schwinger-Keldysh formalism for a
nonequilibrium scalar field. We derive exact time evolution equations
starting from the Schwinger-Dyson equation for the two-point function and
use the three-loop $2PI$ effective action to find a consistent 
approximation to the self energy.
In Sec.~\ref{sec3} we discuss the initial conditions used in this paper.
In order to make contact with heavy-ion physics we consider as initial
state a so-called ``tsunami'': a particle number distribution peaked in
momentum space. Such an initial state is reminiscent of two colliding
high energy wave packets \cite{Pisarski:1997cp}. We hasten to say that in
our $(1+1)$--dimensional scalar field model direct applicability to
phenomology is of course limited. We study in this paper in particular
the time evolution of the Wigner transformed spectral function suitable
for comparison with standard frameworks for kinetic equations. Details of
the Wigner transform are given in Sec.~\ref{sec4}. We present our
numerical results in Sec.~\ref{sec5}. In Sec.~\ref{sec6} we make a
connection with equal-time methods and moment expansions employed before.
Our conclusions are summarized in Sec.~\ref{sec7}.

\section{Time evolution equations} 
\setcounter{equation}{0} 
\label{sec2}

{\it Spectral function.}
We consider a real, scalar quantum field theory with field operators 
obeying the Heisenberg equation of motion
$(\square_x+m^2)\phi(x)+\lambda\phi^3(x)/6=0$.  
We are interested in the nonequilibrium time evolution of the
spectral function $\rho$ given by the expectation value
of the commutator of two fields 
\beq
\rho(x,y)= i\bra [\phi(x),\phi(y)]_-\ket \, .
\eeq
\begin{figure}[t]
\centerline{\psfig{figure=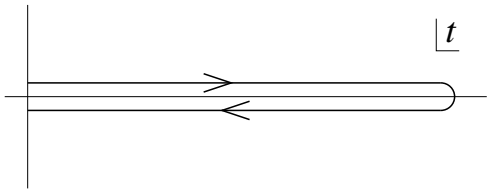,height=2.5cm}}
\caption{Keldysh contour ${\cal C}$ in the complex-time plane.
}
\label{fig1}           
\end{figure}
where $\bra\cdot\ket$ denotes the average over an initial density 
matrix. Nonequilibrium initial value problems can be formulated using the 
Schwinger-Keldysh or closed-time-path formalism
\cite{Schwinger:1961qe}. For a detailed discussion we refer to
Ref.~\cite{Calzetta:1988cq}. In this formalism the theory is formulated
along a contour ${\cal C}$ in the complex-time plane, as shown in
Fig.~\ref{fig1}. Expectation values are defined along this contour. The
two-point function is usually written as
\be
\label{eqG}
G(x,y) = G_>(x,y)\Theta_{\C}(x^0-y^0)+ G_<(x,y) \Theta_{\C}(y^0-x^0), 
\ee
with $\Theta$-functions defined along the closed time path $\C$ 
and the Wightman functions defined as $G_>(x,y)=\bra
\phi(x)\phi(y)\ket = G_<(y,x)$.\footnote{These two-point functions were
also used in Ref.\ \cite{Berges:2000ur}.} For a real scalar field
${G_>}^{\!\!\!\! *}(x,y)=G_<(x,y)$. In terms of these, the spectral
function is given by
\be
\rho(x,y)= i[G_>(x,y)-G_{<}(x,y)].
\ee
The spectral function can be obtained from the propagator
(\ref{eqG}) directly by writing
\be
G(x,y) = F(x,y)-\frac{i}{2} \rho(x,y) \left[ \Theta_{\C}(x^0-y^0)-
\Theta_{\C}(y^0-x^0) \right].
\label{GFR}
\ee
Here we have separated the propagator using the
commutator of two fields, $\rho(x,y)$, and the anti-commutator of two
fields, the symmetric or statistical two-point function
\beq
F(x,y) = \half\bra [\phi(x),\phi(y)]_+\ket
=\half[G_>(x,y)+G_{<}(x,y)].
\label{symF}
\eeq
For a real scalar field  $F(x,y)$ and $\rho(x,y)$ are
real functions with symmetry properties $F(x,y)=F(y,x)$ and
$\rho(x,y)=-\rho(y,x)$. We use $\rho$ and $F$ as the basic quantities in
this paper.

{\it Evolution equations.}
The full propagator (\ref{GFR}) obeys the  
Schwinger--Dyson equation 
\beq
G^{-1}(x,y) = G_{0}^{-1}(x,y) - \Sigma(x,y)
\label{SchwingerDyson}
\eeq
with the proper self energy $\Sigma$ and 
$G_0^{-1}(x,y) = i\, [\square_x + m^2 ]\, \delta^{d+1}_{\C}(x-y)$. 
A differential 
equation for $G$ which may be used to study the time evolution for given 
initial conditions can be obtained from (\ref{SchwingerDyson})
using $\int_{\C} {d}^{d+1}z\, G^{-1}(x,z)G(z,y)
= \delta_{\C}^{d+1}(x-y)$ and one finds  
\beq
\left[ \square_x + M^2(x) \right] G(x,y) 
+\, i \int_{\C}\! {d}^{d+1}z\, \Sigma^{(\rm nonlocal)}(x,z) G(z,y)
= - i \delta_{\C}^{d+1}(x-y) \, .  
\label{GEvolEq}
\eeq
Here we have separated $\Sigma$ in a local part and a nonlocal part,
\beq
\Sigma(x,y) = - i\, \Sigma^{(\rm local)}(x)\, \delta_{\C}^{d+1}(x-y)
+ \Sigma^{(\rm nonlocal)}(x,y) \, .
\label{sighominh}
\eeq
The local part $\Sigma^{(\rm local)}$ corresponds to a mass shift and we
find it convenient to write 
$M^2(x) = m^2 + \Sigma^{(\rm local)}(x)$.

Since our interest focuses on the nonequilibrium dynamics 
of $\rho$ we rewrite (\ref{GEvolEq}) such that the
spectral function appears explicitly as one of the dynamical degrees of
freedom. Similar to the above decomposition for $G$ we can decompose 
the self energy 
\beq
\Sigma^{\rm (nonlocal)}(x,y)=\Sigma_{F}(x,y)
-\frac{i}{2}\, \Sigma_{\rho}(x,y) \left[ \Theta_{\C}(x^0-y^0)-
\Theta_{\C}(y^0-x^0) \right] \, 
\eeq
into two real functions, the statistical component of the self energy
obeying $\Sigma_{F}(x,y)=\Sigma_{F}(y,x)$ and the spectral component
$\Sigma_{\rho}(x,y)=-\Sigma_{\rho}(y,x)$.
These self energies are related to the usual $\Sigma_{>,<}$ as
\bean
\Sigma_\rho(x,y) &=& i[\Sigma_>(x,y)-\Sigma_{<}(x,y)],\\
\Sigma_F(x,y) &=& \half[\Sigma_>(x,y)+\Sigma_{<}(x,y)].
\eean
We also note that the retarded and advanced propagators and self energies
are related to the spectral components as
$G_R(x,y)=\Theta(x^0-y^0)\rho(x,y) = G_A(y,x)$ and $\Sigma_R(x,y) =
\Theta(x^0-y^0)\Sigma_\rho(x,y)=\Sigma_A(y,x)$.

It is straightforward to express (\ref{GEvolEq}) in terms of the real
functions introduced above. The resulting equations are 
\bea
\left[\square_x +M^2(x)\right]\rho(x,y) &=&
-\int_{y^0}^{x^0}\!\!\! dz^0 \int \! d\vecz\,\,
\Sigma_\rho(x,z)\rho(z,y),
\label{eqrho3}
\\
\left[\square_x +M^2(x)\right]F(x,y)\! &=&
- \int_0^{x^0}\!\!\! dz^0  
\int \! d\vecz\,\, \Sigma_\rho(x,z)F(z,y) \nonumber \\
&&+ \int_0^{y^0}\!\!\! dz^0 \int \!d\vecz\,\, \Sigma_F(x,z)\rho(z,y).
\label{eF3}
\eea
The form of these equations is exact and they are equivalent to the standard
Schwinger-Dyson identity (\ref{SchwingerDyson}).     
The equations for $\rho$ and $F$ are explicitly real and causal.
A generic feature is the presence of nonlocal ``memory'' integrals. 
The time integral in Eq.\ (\ref{eF3}) starts at some initial time
taken to be $z^0= 0$. We note that due to the canonical 
commutation relations the
spectral function obeys the 
equal-time properties 
\beq
\label{eqrhoprop}
\rho(x,y)|_{x^0=y^0} = 0, \;\;\;\;
\partial_{x^0}\rho(x,y)|_{x^0=y^0} = \delta^d(\vecx-\vecy).
\label{rhoeqtime}
\eeq
The time evolution of the
spectral function is completely determined by (\ref{eqrho3})
and (\ref{rhoeqtime}).

{\it Three-loop $2PI$ effective action.}
To solve the Eqs.\ (\ref{eqrho3}) and (\ref{eF3}) one has to
find suitable approximation schemes for the self energies
$\Sigma_F$ and $\Sigma_{\rho}$ which respect all symmetries
and allow to describe the nonequilibrium early-time behavior
as well as the large-time approach to equilibrium.
Time reversal symmetry can be easily guaranteed by deriving the
evolution equations from an effective action by a variational 
principle. A systematic way to
achieve the latter is by using the loop expansion of the $2PI$ 
effective action \cite{Cornwall:1974vz}, formulated along the 
closed time path \cite{Chou:1985es,Calzetta:1988cq}. 
Recently, it has been shown that the equations in
the three-loop approximation describe
both the early-time behavior and the physics of 
thermalization \cite{Berges:2000ur,AartsBerges:2001}.
The self energies in this
approximation \cite{Calzetta:1988cq,Berges:2000ur}, here
expressed in terms of $\rho$ and $F$, are for a Gaussian
initial density matrix given by 
\bea
\label{eqSigmaR}
\Sigma_\rho(x,z) &=& 
-\frac{\lambda^2}{2}\rho(x,z)\left[F^2(x,z) -\frac{1}{12}\rho^2(x,z)\right],\\
\label{eqSigmaF}
\Sigma_F(x,z) &=&
-\frac{\lambda^2}{6}F(x,z)\left[F^2(x,z) -\frac{3}{4}\rho^2(x,z)\right]
\eea
(cf.\ Ref.\ \cite{Berges:2000ur} for the extension to quartic initial 
density matrices). The effective
mass term including the local part of the self energy reads 
\beq
M^2(x) = m^2 + \frac{\lambda}{2}\, F(x,x).
\label{effmass}
\eeq 
Eqs.\ (\ref{eqrho3}, \ref{eF3}) with (\ref{eqSigmaR}--\ref{effmass}) 
form a closed set for $\rho$ and $F$ that is energy conserving and
time-reversal invariant.
We emphasize that the propagators in the ``tadpole'' diagram
contribution in Eq.\ (\ref{effmass}) and in the ``setting-sun'' 
contributions (\ref{eqSigmaR}, \ref{eqSigmaF}) are full ones.
The form of the self energies presented here (with free
instead of full propagators) has
also been found in a perturbative calculation in thermal equilibrium
\cite{Aarts:1998kp}.
We note from Eqs.\ (\ref{eqrho3}, \ref{eF3}) that in local approximations,
such as the Hartree approximation, derivable from the two-loop $2PI$
effective action, and leading-order large $N$ schemes,
the evolution of the spectral function $\rho$ does not couple back to the
evolution of the symmetric two-point function $F$.

{\it Thermal equilibrium.} It is instructive to consider the introduced
quantities in thermal equilibrium.  The two-point function $G$ in
equilibrium is described by the same equation (\ref{GEvolEq}) if the
closed time path is replaced by an imaginary path $\C=[0,-i\beta]$, with
$\beta$ the inverse temperature \cite{KadanoffBaym}.  Since the
equilibrium two-point functions depend only on the relative coordinates
it is convenient to consider the Fourier transforms $\rho^{\rm
(eq)}(\omega,\bp)$, etc. From the periodicity (``KMS'') condition for
$G(x,y)$ and $\Sigma^{\rm (nonlocal)}(x,y)$ in imaginary time one finds
the generic equilibrium relations relating for example $G_>^{\rm
(eq)}(\omega,\bp)$ and $G_<^{\rm (eq)}(\omega,\bp)$ \cite{LeBellac}:
\be
G_>^{\rm (eq)}(\omega,\bp) = e^{\beta \om} G_<^{\rm (eq)}(\omega,\bp).
\ee
For the spectral and statistical components we employ here this
translates into the following equilibrium relations
\cite{Aarts:1998kp} 
\bea
F^{\rm (eq)}(\omega,\bp)&=&
-i\Big(n_{\rm B}(\omega)+\frac{1}{2} \Big) \, \rho^{\rm (eq)}(\omega,\bp) \, ,
\\
\Sigma^{\rm (eq)}_F(\omega,\bp)&=&
-i\Big(n_{\rm B}(\omega)+\frac{1}{2} \Big) 
\, \Sigma^{\rm (eq)}_{\rho}(\omega,\bp)
\eea
with $n_{\rm B}(\omega)=(e^{\beta \omega}-1)^{-1}$. While the spectral 
function $\rho^{\rm (eq)}$ encodes the information about the spectrum,
one observes that the symmetric function $F^{\rm (eq)}$ 
encodes the statistical aspects in terms of the particle distribution
function $n_{\rm B}$. For a vanishing $\omega$-dependence the function 
$\Gamma^{\rm (eq)}(\omega,\bp)\equiv \Sigma^{\rm (eq)}_{\rho}(\omega,\bp)/2
\omega$ plays the role of a decay rate for one-particle excited states with
momentum~$\bp$. 

\section{Initial conditions}
\setcounter{equation}{0} 
\label{sec3}

We consider a spatially homogeneous system
such that, after spatial Fourier transformation, the two-point functions
depend only on $\vecp$ and we consider the modes $\rho(x^0,y^0;\vecp)$. 
As indicated above, the initial conditions for $\rho(x^0,y^0;\vecp)$ are fixed
by the equal-time relations (\ref{rhoeqtime}).  As initial conditions for
the symmetric two-point function $F(x^0,y^0;\vecp)$ we consider 
a ``tsunami''--like situation \cite{Pisarski:1997cp} in which a 
spatially homogeneous collection of particles  
move with approximately the same momentum 
peaked around 
$\vecp_{\rm ts}$ and $ - \vecp_{\rm ts}$. Explicitly, we 
take
\bea
\nonumber
F_0(x^0,y^0;\vecp)|_{x^0=y^0=0} &=& 
\frac{1}{\om_\vecp}\left[n_0(\vecp)+\half\right],\\
\partial_{x^0}  \partial_{y^0} F_0(x^0,y^0;\vecp)|_{x^0=y^0=0} &=&
\om_\vecp\left[n_0(\vecp)+\half\right],\\
\partial_{x^0}F_0(x^0,y^0;\vecp)|_{x^0=y^0=0} &=& 0,
\nonumber
\eea
with $\om_\vecp = \sqrt{\bp^2 + m^2}$ and the initial particle number
\be
n_0(\vecp) = {\cal N}\exp\left(-\frac{1}{2\sigma^2}
(|\vecp|-|\vecp_{\rm ts}|)^2\right).
\label{initialcondition}
\eeq
Here $\sigma$ controls the width of the initial distribution and 
${\cal N}$ is a normalization constant. 
The initial condition is clearly far from thermal equilibrium and
reminiscent of two colliding high energy wave packets 
\cite{Pisarski:1997cp}.

To study the nonequilibrium evolution of the spectral function
$\rho(x^0,y^0;\vecp)$ we solve the coupled Eqs.\ 
(\ref{eqrho3}--\ref{eF3}) with (\ref{eqSigmaR}--\ref{effmass})
numerically along the lines of ref.\ \cite{Berges:2000ur}. 
We use a standard discretization on a lattice in space and time for
a finite spatial volume and spatially periodic boundary conditions. To
remove finite size effects we increase the volume until the results
become stable.   
Because the evaluation of the memory integrals is rather time- and
memory-consuming the numerical results presented below are for a
$(1+1)$--dimensional system.
The lattice acts as a cutoff and regulates the ultraviolet divergences, 
which have to be renormalized. In $1+1$ dimensions this is straightforward
since only the one-loop contribution is logarithmically
divergent. We absorb this in a bare mass parameter $\mu$ with the
replacement $m^2 \mapsto \mu^2 = m^2-\delta m^2$ in Eq.\ (\ref{effmass}). 
The counterterm $\delta m^2$ cancels the divergent vacuum contribution
coming from the one-loop graph. The finite part of $\delta m^2$ is fixed by
requiring that the renormalized one-loop mass parameter in vacuum 
($n_0(\vecp)\equiv 0$) equals $m$ and we express all dimensionful scales
in units of $m$. For the plots we have used a space lattice with spacing 
$ma=0.3$ and a time lattice $a_0/a=0.25$. The system size is $mL=24$.

\section{Wigner transformation}
\setcounter{equation}{0} 
\label{sec4}

Derivations of transport or kinetic equations are typically performed
by considering correlation functions in the so-called Wigner representation. 
Two-point functions are
written in terms of the center-of-mass coordinate $X=(x+y)/2$ and the relative
coordinate $x-y$, with respect to which a Fourier transformation to momentum
space is performed (Wigner transformation). Derivations based on a 
gradient expansion in $\partial_X$ assume a slow variation of the
two-point functions with $X$. In particular, the classical Boltzmann equation
can be obtained from (\ref{eqrho3}--\ref{effmass}) using a first-order
gradient expansion and an on-shell form of the spectral function with
zero width \cite{Calzetta:1988cq}. The full equations  
(\ref{eqrho3}--\ref{effmass}) can therefore be considered as a
``quantum Boltzmann equation'' including off-shell effects and resumming 
an infinite order of derivatives.  

To analyze the evolution of the spectral function (not to solve the
dynamics) we perform a Wigner transformation and write
\be
\label{eqW}
i\rho(X^0; \om, \vecp) = \int_{-2X^0}^{2X^0}dt\, e^{i\om t}
\rho(X^0+t/2, X^0-t/2;\vecp).
\ee
The $i$ is introduced such that $\rho(X^0;\om,\vecp)$ is real. Because the 
spectral function is antisymmetric, $\rho(X^0;\om, \vecp) = 
-\rho(X^0;-\om, \vecp)$, we will present the positive-frequency part
only.
Furthermore, as a consequence of the equal-time relation (\ref{eqrhoprop})
the Wigner transform obeys the sum rule $I = \int d\om/(2\pi)\, \om\,
\rho(X^0;\om,\bp) = 1$. In our simulations an evaluation of the sum
rule typically results in $1.000<I<1.003$, indicating the accuracy in the
numerical evaluation of the Wigner transform.

Since we consider an initial-value problem with  
$x^0,y^0\geq 0\,$, the time integral over $t=x^0-y^0$ is bounded by
$\pm 2X^0$. This leads to a so-called ``finite-time effect'' in the
Wigner transformed quantities. To understand the qualitative aspects 
of a finite-time interval it is instructive to consider the free spectral 
function  $\rho_0(x^0,y^0;\vecp) = 
\om_\vecp^{-1}\, \sin [ \om_\vecp (x^0-y^0) ]$ with 
$\om_\vecp = \sqrt{\bp^2 + m^2}$.  After performing a Wigner
transformation, we find
\be
\label{eqW0}
\rho_0(X^0;\om,\bp) = 
\frac{\sin [(\om-\om_\vecp)2X^0]}{\om_\vecp(\om-\om_\vecp)} -
\frac{\sin [(\om+\om_\vecp)2X^0]}{\om_\vecp(\om+\om_\vecp)} \, . 
\ee 
For finite $X^0$ this spectral function shows a rapidly oscillating
behavior, while its envelope is peaked at $\om=\pm\om_\vecp$.
In the limit $X^0\to \infty$ it reduces to $\rho_0(\om,\vecp) =
2\pi\mbox{sgn}(\om)\delta(\om^2-\om_\vecp^2)$. 

For an interacting system, however, effective damping leads to a finite
correlation time $\tau$, such that finite-time effects are expected
to vanish when $X^0\gg\tau$. This can be investigated
analytically by assuming
that the spectral function is strictly exponentially damped and can be
written as $\rho(x^0,y^0;\vecp)=e^{-\gm_\vecp|x^0-y^0|}E_\vecp^{-1}\sin
[E_\vecp(x^0-y^0)]$. The effective mass $E_\vecp$ and rate $\gamma_\vecp$
are allowed to depend on the time $X^0$.  The corresponding Wigner
transform reads $\rho(X^0;\om,\vecp) = \rho_{\rm BW}(X^0;\om,\vecp) + 
\delta\rho(X^0;\om,\vecp)$, where $\rho_{\rm BW}$ denotes the 
Breit-Wigner function
\be
\rho_{\rm BW}(X^0;\om,\vecp) =
\frac{2\om\Gamma_\vecp(X^0)}{[\om^2-E_\vecp^2(X^0)]^2
+ \om^2\Gamma^2_\vecp(X^0)}
\label{BreitWigner}
\ee
with a width $\Gm_\vecp(X^0)=2\gm_\vecp(X^0)$. The additional contribution
$\delta\rho(X^0;\om,\vecp)$ vanishes exponentially as $\exp(-\Gamma_\vecp
X^0)$, such that the finite-time effect is indeed absent for large enough
(but finite) $X^0$. We want to emphasize that this finite-time effect is a
general consequence of an initial-time problem formulated in terms of
Wigner transformed quantities and should be taken into account when used to
describe the early-time behaviour ($\Gamma_\vecp X^0 \lesssim 1$) in an
experimental
setup.

Finally, we note that the positivity condition,
$\mbox{sgn}(\om)\rho(X^0;\om, \vecp)\geq 0$, can only be shown to hold in
the special case that the initial density matrix commutes with the full
Hamiltonian, such as in thermal equilibrium. In this case the system is of
course stationary and independent of $X^0$.  As a consequence the
interpretation of the nonequilibrium spectral function as the density of
states should be taken with care.

\section{Nonequilibrium evolution}
\setcounter{equation}{0} 
\label{sec5}

\begin{figure}[t]
\centerline{\psfig{figure=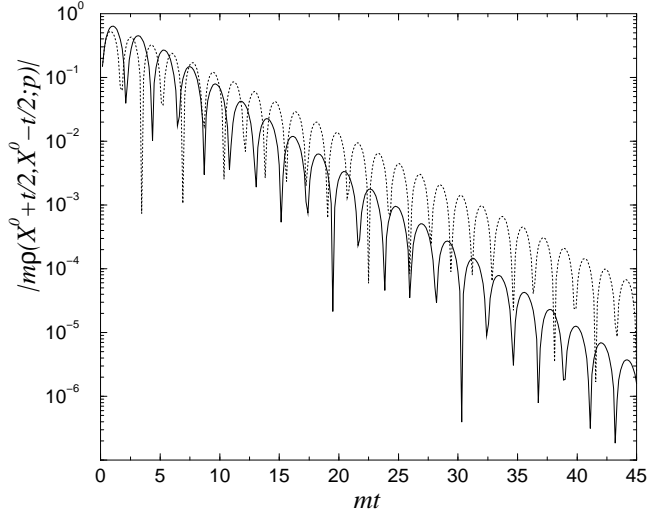,height=7cm}}
\caption{Absolute value of $\rho(X^0/2+t,X^0/2-t;p)$ as a function of $t$
at fixed $mX^0=35.1$ for spatial momentum $p=2\pi
k/L$ with $k=0$ (full), $4$ (dotted). 
The coupling is $\lambda/m^2=4$. 
}
\label{fig33_1}           
\end{figure}

In Fig.\ \ref{fig33_1} we show the time evolution of the spectral function
$\rho(X^0+t/2,X^0-t/2;p)$ as a function of the relative time $t$ for two
different momentum modes at given $X^0$, as obtained from the
numerical solution of Eqs.\ (\ref{eqrho3}--\ref{effmass}). From the 
logarithmic plot one observes an effective damping of the oscillating 
spectral function while the oscillations never damp out completely.
The decrease of the maximum amplitude very quickly 
approaches an approximately exponential behavior characteristic for the
decay
of one-particle excited states with given momentum. One observes
that the zero momentum mode is stronger damped than the higher
momentum mode.\footnote{An analysis of damping times as
a function of the coupling $\lambda$ can be found in Ref.\ 
\cite{Berges:2000ur}.}

\begin{figure}[t]
\centerline{\psfig{figure=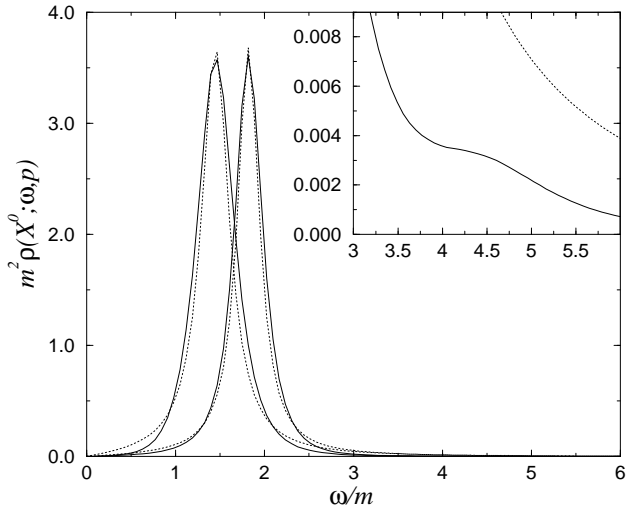,height=7cm}}
\caption{Wigner transforms $\rho(X^0;\om,p)$ of the real-time spectral
functions of Fig.\ \ref{fig33_1} as a function of $\om$
at $mX^0=35.1$ for $k=0$ (left peak) and $k=4$ (right peak). 
Also shown are fits to a Breit-Wigner function (dotted) with 
$(E_{p}/m,\Gamma_{p}/m)= (1.46,0.37)$ ($k=0$) and
$(1.82,0.30)$ ($k=4$). 
The inset shows a blow-up of the zero mode around the 
three-particle threshold $3E_{p}/m=4.38$. 
The expected bump from off-shell decay is small but visible.
}
\label{fig33_4}           
\end{figure}

The effective damping should correspond to a nonzero width of the Wigner
transformed spectral function. In Fig.\ \ref{fig33_4} we display the Wigner
transforms $\rho(X^0;\om,p)$ for the zero momentum mode (left peak) and the
higher momentum mode (right peak) for the same $X^0$-time as in
Fig.~\ref{fig33_1}.  One clearly observes that the interacting theory has a
continuous spectrum described by a peaked spectral function with a nonzero
width. The inset shows a blow-up of the zero mode around the three-particle
threshold $3E_{p}/m=4.38$. The expected bump in the spectral function is
small but visible. We stress that this bump in the spectral function is
kinematically forbidden for the on-shell limit and arises purely from
off-shell decay. 
In Fig.\ \ref{fig33_4} we also present fits to a Breit-Wigner spectral
function. While the position of the peak can be fitted
easily, the overall shape and width are only qualitatively captured.
In particular, the slope of $\rho(X^0;\om,p)$ for small $\om$ is 
quantitatively different. We also see that the Breit-Wigner fits give
a narrower spectral function (smaller width) and therefore would predict a
slower exponential relaxation in real time.

\begin{figure}[t]
\centerline{\psfig{figure=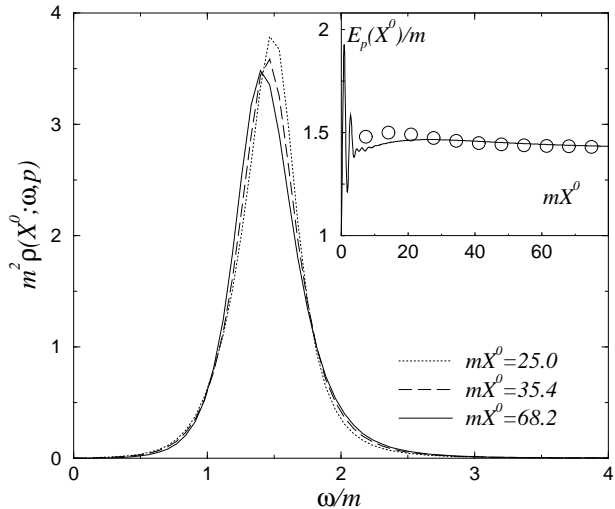,height=7cm}}
\caption{Time dependence of the spectral function $\rho(X^0;\om,p)$ at
three values of $X^0$ for $p=0$. The peak slowly shifts towards lower
values. 
The inset shows the $X^0$-dependence of $E_{p}(X^0)$ obtained 
from a fit to a Breit-Wigner function (circles) 
as well as the one-particle energy $\ep_p(X^0)$ defined via 
equal-time two-point functions, see Eq.\ (\ref{Eonepart}) below (line).
}
\label{fig33_15}           
\end{figure}

We now turn to the dependence of $\rho(X^0;\om,p)$ on $X^0$.  In Fig.\
\ref{fig33_15} we show the spectral function at zero momentum
$\rho(X^0;\om,p=0)$ for three different values of $X^0$. We see that the
peak slowly shifts to smaller values. In the inset to Fig.\ \ref{fig33_15}
this behavior is quantified by plotting the $X^0$-dependence of
$E_{p}(X^0)$ obtained from a fit to the Breit-Wigner function
(\ref{BreitWigner}). For comparison, the solid line in the same figure
shows the time-dependent one-particle energy $\epsilon_{p}(X^0)$ as
defined
via equal-time two-point functions in Eq.\ (\ref{Eonepart}) below. We see a
good agreement of both definitions of the one-particle energy for large
enough times. We come back to this below.

\begin{figure}[t]
\centerline{\psfig{figure=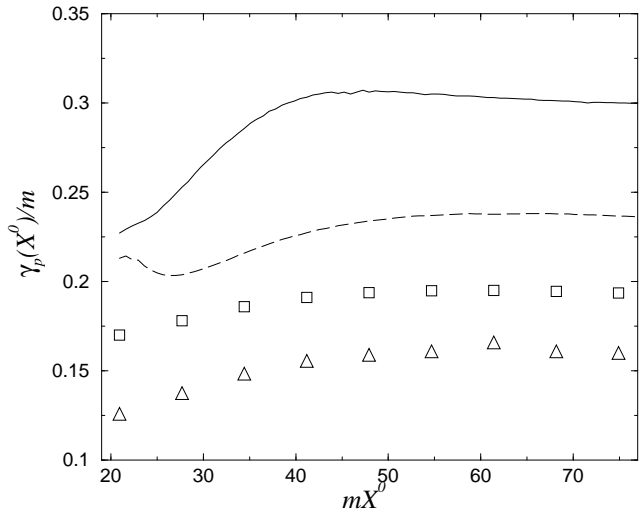,height=7cm}}
\caption{Time dependence of the relaxation
rate $\gamma_{p}(X^0)$, determined directly in real time from 
$\rho(X^0+t/2,X^0-t/2;p)$ for $p=2\pi k/L$,
$k=0$ (full) and $k=4$ (dashed).    
Also shown is $\Gamma_{p}(X^0)/2$ from a fit to a Breit-Wigner function, 
for $k=0$ (squares) and $k=4$ (triangles). 
}
\label{fig33_13}           
\end{figure}

A qualitative way to characterize the $X^0$-dependence of the shape of the
spectral function is to assume that it is approximately described by a
Breit-Wigner function.  The time dependence of $\Gamma_p(X^0)$ can then be
followed by performing fits for different values of $X^0$. The resulting
$\Gamma_p(X^0)/2$ is shown in Fig.\ \ref{fig33_13} for two momentum modes.
After a variation for early times, the $X^0$-dependence becomes weaker for
larger times.  As indicated before (see Fig.\ \ref{fig33_4}), the
Breit-Wigner fits result in a smaller width than what is expected by
comparing with the actual spectral function. A simple way to obtain a
quantitative description is by extracting a relaxation rate $\gamma_p$
directly in real time, i.e.\ from a fit of the spectral function
$\rho(X^0+t/2,X^0-t/2;p)$ to an exponential form $\sim
\exp[-\gamma_p(X^0)t]$ (see Fig \ref{fig33_1}). The resulting rates are
also presented in Fig.\ \ref{fig33_13}. We observe that the relaxation rate
for the zero mode increases first, corresponding to a broadening of the
Wigner transformed spectral function, before it levels off at later times.  
Despite the quantitative differences we see a similar qualitative
$X^0$-dependence from the Breit-Wigner fit for large enough times.

{\it Particle distribution function.}
The relatively weak $X^0$-dependence of the kinetic aspects of the
evolution encoded in the spectral function has to be confronted 
with the time dependence of the statistical aspects encoded in $F$.
The particle distribution is expected to exhibit strong qualitative
changes,    
since we start with an initial condition (\ref{initialcondition})
far away from thermal equilibrium and the current approximation
is known to describe the physics of thermalization at large enough 
times \cite{Berges:2000ur}.

\begin{figure}
\centerline{\psfig{figure=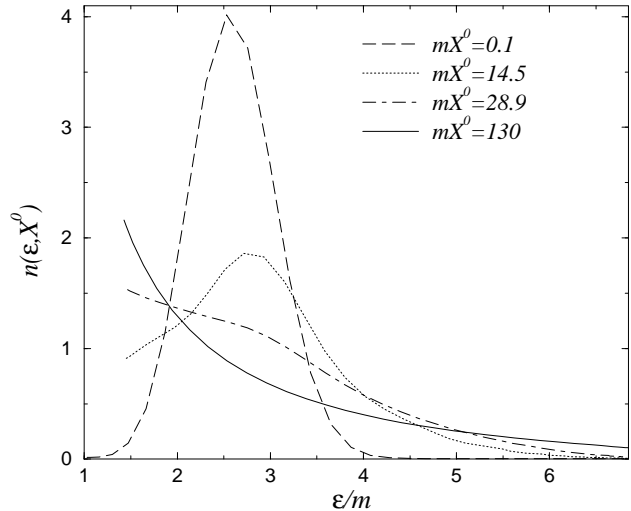,height=7cm}}
\caption{Effective particle number $n(\ep_p, X^0)$ as a function of
$\ep_p$ at four values of $X^0$.
As time goes on, the original ``tsunami'' vanishes and the infrared modes
become more excited.   
}
\label{fig33_7}           
\end{figure}

The evolution of the statistical information can be extracted
from the equal-time correlation function $F(t,t;p)$. 
If the theory exhibits a quasi-particle
structure it is possible to define a quasi-particle distribution function 
from a combination of equal-time two-point functions
\cite{Aarts:2000zn,Salle:2000jb}. For this purpose we introduce the
two-point function $K(x,y)=\half\bra [\pi(x),\pi(y)]_+\ket$, where
$\pi(x)=\partial_{x^0}\phi(x)$ is the
canonical momentum. In the spatially homogeneous situation we consider here,
$K(x^0,y^0;p)=\partial_{x^0}\partial_{y^0}F(x^0,y^0;p)$. 
We define the distribution function $n(\ep_p(X^0), X^0)$ at time $X^0$ as
\be 
n(\ep_p, X^0)+\frac{1}{2} =
\left[K(X^0,X^0;p)F(X^0,X^0;p)\right]^{1/2} \, .
\label{distfunc}
\ee
For the free theory the definition of $n(\ep_p, X^0)$ corresponds
to the standard one-particle distribution. The distribution function 
depends on a time\--dep\-en\-dent one-particle energy, defined as 
\be
\ep_p(X^0) = 
\left[K(X^0,X^0;p)/F(X^0,X^0;p)\right]^{1/2}.
\label{Eonepart}
\ee
The applicability of this definition to describe one-particle energies
for the present nonequilibrium situation finds
an impressive demonstration by comparing it with the position of the peak in
the spectral function as discussed above (cf.\ Fig.\ \ref{fig33_15}).

In Fig.\ \ref{fig33_7} we show the results for the effective particle
number $n(\ep_p, X^0)$ as a function of $\ep_p$ at four values of $X^0$.
The initially high occupation number in a small momentum range rearranges
as time goes on. The original ``tsunami'' vanishes and the infrared modes
become more excited. The distribution is not yet thermal. Thermalization
takes place on longer time scales which is shown for this model in Ref.\
\cite{Berges:2000ur}. The strong qualitative changes of the statistical
aspects encoded in the particle distribution function have to be confronted
with the comparably moderate changes of the spectral function shown in
Fig.\ \ref{fig33_15} and \ref{fig33_13}.

\section{Equal-time spectral function and moments}
\setcounter{equation}{0} 
\label{sec6}

In order to make a connection with methods employed before, 
we discuss some equal-time properties of the spectral function.
As has been discussed earlier, the spectral function is the commutator of
fields and obeys elementary equal-time properties (\ref{rhoeqtime}).
There is a further hierarchy of equal-time relations that
follow from the exact evolution equation (\ref{eqrho3}).
The identity
\be
\label{eqd2}
\partial^2_{x^0}\rho(x^0,y^0;\vecp)|_{x^0=y^0} = 0
\ee
can also be traced back directly to the operator equation of motion.
The following three equal-time derivatives are less trivial and read
\bea
\nonumber
\partial^3_{x^0}\rho(x^0,y^0;\vecp)|_{x^0=y^0} &=&
-\left[\vecp^2+M^2(x)\right],\\[0.15cm]
\label{eqequaltime}
\partial^4_{x^0}\rho(x^0,y^0;\vecp)|_{x^0=y^0} &=&
-2\partial_{x^0}M^2(x),
\\[0.15cm]
\nonumber
\partial^5_{x^0}\rho(x^0,y^0;\vecp)|_{x^0=y^0} &=&
\left[\vecp^2+M^2(x)\right]^2 - 3\partial^2_{x^0}M^2(x)\\
\nonumber
&&-\partial_{x^0}\Sigma_\rho(x^0,y^0;\vecp)|_{x^0=y^0}.
\eea
These expressions are exact. One observes that the self energy 
$\Sigma_\rho$, necessary to describe scattering, enters for the first time
only in the fifth derivative.
This suggests that in formalisms which are strictly based on 
equal-time correlation functions \cite{Wetterich:1997ap}, the kinetic 
aspects of scattering encoded in the spectral function may only be 
captured in very sophisticated truncations.

The above relations can be immediately recast in terms of moments of the
spectral function in the Wigner presentation. We define the $n$'th moment
as
\be
A_n(X^0;\vecp) = \int_{-\infty}^{\infty}\frac{d\om}{2\pi}\,\om^n
\rho(X^0;\om,\vecp).
\ee
Due to the antisymmetry of $\rho(X^0;\om,\vecp)$ all even moments 
$A_{2n}(X^0;\vecp)$ vanish. It is straightforward to relate the odd
moments directly to the equal-time derivatives given above, using
$\partial_{x^0} = \half \partial_{X^0} + \partial_{t}$. The first moment is
just the sum rule, $A_1(X^0;\vecp)=1$ and Eq.\ (\ref{eqd2}) provides a
consistency check, $\partial^2_{x^0}\rho(x^0,y^0;\vecp)|_{x^0=y^0} =
\partial_{X^0}A_1(X^0;\vecp) = 0$. For the higher derivatives one finds
\bea
\nonumber
\partial^3_{x^0}\rho(x^0,y^0;\vecp)|_{x^0=y^0} &=&-A_3(X^0;\vecp),\\[0.15cm]
\label{eqmom}
\partial^4_{x^0}\rho(x^0,y^0;\vecp)|_{x^0=y^0} &=&
-2\partial_{X^0}A_3(X^0;\vecp),\\
\nonumber
\partial^5_{x^0}\rho(x^0,y^0;\vecp)|_{x^0=y^0} &=&
-\frac{5}{2}\partial^2_{X^0}A_3(X^0;\vecp) +A_5(X^0;\vecp).
\eea
The combination of (\ref{eqmom}) and (\ref{eqequaltime}) gives information
on the time dependence of the lowest moments. Therefore, also in
approximation schemes that are based on an expansion in moments, the lowest
quantity that shows a direct dependence on the self energy $\Sigma_\rho$
appears only at relatively high (fifth) order. We conclude that both in
equal-time formalisms and in approximations based on moments a proper
inclusion of scattering effects might be difficult to achieve.

\section{Conclusion}
\label{sec7}

The presence of a nonequilibrium environment affects the quasiparticle
structure of the theory, encoded in the spectral function.
As a relevant physical application one may think of high energy
heavy-ion collisions where spectral properties of the vector mesons
are affected by the presence of nuclear matter.

In this paper we have concentrated on a self-consistent determination of
the spectral function out of equilibrium, using a scalar field theory in
$1+1$ dimensions as our (toy) model.  We have calculated the
nonequilibrium time evolution of the spectral function by solving the
equations of motion obtained from the three-loop $2PI$ effective action
in real time. To compare the dynamics of the spectral function and the
statistical aspects of the theory we also compute the evolution of an
effective particle number distribution.  Scattering and off-shell
contributions are included in the dynamical evolution, in contrast to
well-studied Hartree and leading-order large $N$ approximations. Even for
moderate couplings and initial conditions far away from equilibrium, we
observe a rather weak dependence of the Wigner transformed spectral
function on $X^0$. This fact provides a necessary condition for a
successful gradient expansion typically employed in the derivation of
kinetic equations. After variations at early times we observe that the
spectral function approaches a rather stable shape characterized by a
nonvanishing width. A consistent ``quantum Boltzmann equation'' certainly
has to relax the zero-width assumption employed in the derivation of the
standard Boltzmann equation. We have argued that it might be
notoriously difficult to derive the kinetic aspects of scattering encoded
in the spectral function from equal-time formalisms or approximation
schemes based on an expansion in moments. We emphasize that the employed
framework based on the $2PI$ effective action allows for a consistent
description of both the early and the late time behavior.
Since a change in the position of the peak and width of the
spectral function can be described consistently in this framework, it
would be interesting to extend the study presented here to more realistic
models for heavy-ion collisions.

Finally the analysis performed in this paper was largely motivated by
the physics of relativistic heavy-ion collisions. We would like to
stress, however, that the approach using the $2PI$ effective action is
general and that possible applications of the formalism range from the
early universe to condensed matter systems.

\vspace{5mm}
                 
{\bf Acknowledgements}

\noindent{
J.B.\ thanks J\"urgen Cox for collaboration on related subjects.
We would like to thank Tomislav Prokopec for many 
discussions.
This work was supported by the TMR network {\em Finite
Temperature Phase Transitions in Particle Physics}, EU contract no.\
FMRX-CT97-0122.}

\end{document}